\documentstyle{lmamult}
\input psfig
\begin{document}
\def\simlt{\lower.5ex\hbox{$\; \buildrel < \over \sim \;$}}
\def\simgt{\lower.5ex\hbox{$\; \buildrel > \over \sim \;$}}
\title {REIONIZATION AND THE COSMIC MICROWAVE BACKGROUND\footnote{ To appear in
{\it Present and Future of the Cosmic Microwave Background}, eds. J. L. Sanz,
E. Martinez--Gonzalez and L. Cayon (in press, 1994)}}
\titlerunning{Reionization}

\author {Joseph Silk}

\institute{Departments of Astronomy and Physics, and
Center for Particle Astrophysics, University of California, Berkeley,
CA 94720, USA}

\maketitle


\noindent{\bf Abstract}

\noindent{If the COBE  detection of CMB fluctuations is used to normalize the
power
spectrum of primordial density fluctuations in a cold dark matter--dominated
universe, early reionization is likely to result in a substantial diminution of
primordial temperature fluctuations on degree scales.  I argue that the
reionization may be non--Gaussian because of feedback effects.  Secondary
fluctuations on arc--minute scales provide an important probe of the efficiency
of rescattering.}

\section{Introduction}

We are rightly celebrating at this workshop the triumph of the COBE DMR
experiment in measuring fluctuations on $\simgt 10^\circ$ angular scales in the
cosmic microwave background.  This provides the long--sought normalization for
the primordial density fluctuation spectrum that gave rise via gravitational
instabilities to the present--day structure.  Yet there are many choices that
remain in models for primordial fluctuations as well as in cosmological
parameters.  An angular scale of $10^\circ$ on the last scattering surface
corresponds to a comoving scale of 1 Gpc, useful for establishing the validity
of the cosmological principle, and for probing inflation,  but of little direct
relevance to formation of
known structure.  Only by detecting fluctuations on the precursor scales of the
largest structures (degree scales) and galaxy clusters (arc--minute scales) can
we hope to specify more precisely the elusive models that quantify the details
of galaxy formation.  One important complication arises from the thermal
history of the universe since redshift $z=1000,$ where the last scattering
surface is located in the canonical uniform  model of the expanding universe.

The principle purpose of this talk is to argue for the inevitability of
reionization in the standard model for large-scale structure, that of cold
dark matter, and to discuss, more generally, the secondary signatures of early
reionization for microwave  background anisotropies on small angular scales.
First, however, I describe a new presentation of data from experiments on
different angular scales that largely circumvents any assumptions about the
radiation
power spectrum.

\section{Comparison of Different Experiments}

Microwave background experiments probe a complex domain of sky and frequency
sampling.  To facilitate comparison between  different experiments, a novel
transformation has been developed that transforms the Gaussian autocorrelation
function angular fits into something more physically useful.

\dimen10=\hsize \divide\dimen10 by 1
\setbox10=\hbox{\psfig{file=bunn1.ps,height=12cm,width=\dimen10}}
\vglue -1.3cm
\vskip5truecm
\centerline{$\vcenter{\box10}$}
\vglue 0cm
\setbox10\vbox{\overfullrule=0pt \parindent=0pt \hsize=\dimen10
Figure 1.  Experimental filter-averaged detections and upper limits, compared
to radiation power spectra for CDM $\Omega_B=0.12$ model for various redshifts
of reionization, as labelled (top); experimental filters (bottom) }
\centerline{$\vcenter{\box10}$}
\vskip10pt

The approach,
pioneered by Bunn \cite{Bun} and performed independently by Bond \cite{Bon}, is
straightforward.  The temperature fluctuations are expanded in spherical
harmonics ${\delta T \over T} = \Sigma a_{lm} Y_{lm}$ with power per harmonic
$C_l = <\vert A_{lm} \vert^2>.$  The contribution to the total power per
decade is
$B_l =l(2l+1){C_l \over 4\pi}.$  If the filter function of a particular
experiment that defines its sensitivity to harmonics $l$ is $F_l,$ then one can
define ${\bar B} = {\Sigma B_lF_l \over \Sigma F_l} $ as the experimental
filter function--averaged power.  One can show by computing $\bar B$ for
different input
power spectra that it is insensitive to the assumed power spectrum.

Figure 1 displays a comparison between the experimentally-averaged power for
various recent detections of cosmic microwave background fluctuations and the
radiation power spectra for various cold dark matter models, characterized, in
the spirit of the pre
angular scales is significant, because the uncertainties in modelling out
Galactic foreground contamination have not been fully included, and also
because of the very limited sky coverage. I discuss below a possible
cosmological source of such variations.


\section{Reionization with Cold Dark Matter}

Early nonlinearity will result in production of dust and ionization input that
can reionize the intergalactic medium.  Indeed, this must have happened with
high efficiency prior to redshift $z\approx 5.$  The lack of a Gunn--Peterson
absorption trough towards high redshift quasars means that the neutral fraction
in the intergalactic medium is less than $10^{-6}.$  Indirect pointers to early
energy input come from the enrichment of the intracluster medium, where there
is more iron in the gas than in the stellar components of the cluster galaxies.
 This requires considerable ejection of enriched gas during the early stages
of cluster galaxy evolution.  Ejection  from supernovae would inevitably
involve input of heat into the intergalactic medium.  Another indirect
indicator of early heat and ionization input comes from the excess, relative to
the
present day, population of faint blue dwarf galaxies required to account for
the deep galaxy counts.  The lack of a present day counterpart in the field
requires such galaxies, also a common feature in theoretical models, to have
ejected considerable amounts of their initial mass in a supernova--driven wind.
 In nearby galaxy clusters, there is a population of low surface brightness
dwarfs that may be the relics of these early, star forming dwarfs:  again,
considerable mass loss must presumably have occurred in order to differentiate
 the low surface brightness dwarfs from the compact dwarf galaxies.

If there is so much energy input at $z\simlt 5,$ it is not much of an
extrapolation to imagine that a more modest energy input occurred at an earlier
epoch.  Only $\sim 20\rm eV$ per baryon, a miniscule fraction of the proton
rest mass, is required to photoionize the intergalactic medium.  Once ionized,
the
gas can recombine.  However the ratio of recombination time to expansion time
 is approximately $0.06h^3 ({100/ (1+z)})^{3 \over 2}.$  Hence the required
efficiency of conversion of mass into, say, ionizing photons is
 still low even at $z\sim 100.$

Various structure formation models are capable of early $(z \simgt 50)$
reionization.  These include models with intrinsically nonlinear
objects, such as texture or string--seeded models.  Another class of models in
which early reionization is natural and even inevitable is the baryon dark
matter--dominated universe.  These models are only viable if the primordial
density fluctuations are isocurvature.  The spectrum of isocurvature density
fluctuations imposed by large--scale structure constraints \cite{CenOstPee}
${\delta \rho \over \rho}
\propto M^{-{n+3 \over 6}},$ $n \approx -0.5,$ diverges towards small scales;
$n\simlt 0$ is required observationally to avoid excessive power on Mpc scales,
and $n<1$ guarantees divergence in terms of potential fluctuations on small
scales early in the m
period:  there need be no hydrogen recombination epoch.  The more conventional,
cold dark matter--dominated universe, if normalized to reproduce the COBE DMR
measurement of ${\delta T \over T} \approx 10^{-5}$ on large angular scales,
has early collapse of small, rare objects.  These can, with reasonable
efficiency, reionize as early as $z\sim 100.$

Many models fail to have sufficient power to reionize at an epoch sufficiently
early that some smoothing of CMB fluctuations occurred.  These include mixed
dark matter (30 percent hot, 70 percent), tilted $(n\approx 0.7),$ and
vacuum--dominated cold dark matter $(\Omega_{vac} \approx 0.8)$ models, all
designed to
reconcile COBE $\delta T \over T$ with large--scale $(\sim 1 - 50 h^{-1} \rm
Mpc)$
structure observations.  In a cold dark matter--dominated universe, however,
with mass tracing light and $\Omega =1,$ the typical object to go nonlinear at
$z\sim 30$ has mass $\sim 10^4 \rm M_{\odot}.$

An investigation of reionization in a cold dark matter--dominated universe led
to the following conclusions \cite{TegSilBla}.  The fraction of the rest mass
density of
nonlinear objects that goes into ionizing photons is taken to be a free
parameter.  With one percent efficiency of producing ionizing photons and
unbiased cold dark matter,
ionization is complete at $z=80.$  If $f =0.006,$ as might be appropriate
if 90 percent of the baryons formed massive stars that produced ionizing
photons, reionization occurs as early as $z=130.$  It is difficult to imagine
more efficient conversion of baryons into ionizing photons.By contrast, a
tilted model could not reionize before $z \sim 20.$

If the universe is reionized back to redshift $z,$ the scattering probability
of CMB photons depends only on the combination $\Omega_b h.$  From primordial
nucleosynthesis, $\Omega_bh \approx 0.015,$ so that for a cold dark matter
universe, one might take $\Omega_b h \approx 0.03$ and infer a scattering
probability of $0.7$ at $z=100,$ $0.4$ at $z = 60,$ $0.2$ at $z = 30.$ The
corresponding angular scales subtended at these epochs are $4^\circ,
5^\circ,$ and $6^\circ.$  For $\Omega h = 0.03,$ the probability distribution
for the redshift which a CMB photon was last scattered, or visibility function,
peaks at $z \approx 40.$  For a CMB photon to have been scattered at least once
with probability $0.5$ only requires an efficiency $\sim 10^{-4}$ in a
canonical model after typical low mass objects undergo first collapse at
 $z \sim 40.$

\section{Generation of Secondary Fluctuations}

Reionization at $z\simgt 20$ partially smooths out degree-scale CMB
fluctuations, but regenerates small angular scale fluctuations.  This second
order effect arises because in first order, no fluctuations arise.  To see
this, one studies the photon collision terms that are source terms in the
second order Boltzman equation  for the radiation intensity, $\Delta$
(equivalent to ${\delta \rho_{\gamma}\over \rho_{\gamma}}),$ namely
$${\partial \Delta \over \partial t} + {\gamma_i} {{\partial \Delta} \over
{\partial x_i}} = n_e \sigma_T [1+\delta ({\bf x})]\left\{ -\Delta + 4
{\bf\gamma}{.}{\bf v} -v^2 +\ldots  \right\}.$$

Here $\gamma_i$ is the photon direction, $\delta ({\bf x})$ is the first order
matter fluctuation at position ${\bf x},$ ${\bf v}$ is the matter
velocity.  To first order, the right hand side of the Boltzman equation has a
term $(-\Delta n_e \sigma_T)$ that, arising from primary last scattering,
 is exponentially suppressed by rescattering as exp
$(-\int n_e \sigma_T cdt).$
The Doppler term linear in $v,$ $\sim n_e \sigma_T v ,$ is suppressed by phase
cancellation.  The second order terms survive.  However the $O(v^2)$
and other terms are negligible, and the only significant contribution term is
the $O (\delta v)$ term.  This is known as the Vishniac effect.  Since
it is in the convolution of two terms that, in the BDM model, at most diverge
weakly towards smaller scales when averaged over the thickness of the last
scattering surface $\delta \propto R^{-{n+3\over 2}},$ $ v \propto
R^{-{n+1 \over 2}},$ it diverges significantly, ${\delta T\over T} \propto
\delta v \propto R^{-n-2},$ $ n\approx {-{1 \over 2}},$ and is of
order $10^{-5}$ on subarcminute scales if $\Omega \sim 0.2.$
In reionized CDM models, the Vishniac effect amounts at most to
${\delta T \over T} \sim 10^{-6}$ for any realistic ionization history and
baryon
density.

Since observational limits are about ${\delta T \over T} \sim 10^{-5}$ over an
arcminute, from experiments at  OVRO and ATCA, one can strongly constrain the
allowable $(\Omega,n)$ parameter space for the BDM model \cite{HuScoSil}.  Some
parameter space
remains, near $n \approx -0.5,$ $\Omega \approx 0.1 -0.2,$ where the BDM model
happens to have the parameters that are preferred by large--scale structure
observations.

\section{Reionized CDM}

While the CDM predictions on small scales are well below current limits, the
factor $\simlt 2$ suppression that is likely on degree scales is significant.
If one takes the lowest limits to date on degree scales at face value, then
consistency with the COBE detection on $10^\circ$ and the CDM prediction
for $n=1$ initial conditions has been questioned.  Smoothing makes a
significant difference to this conclusion:  even reionization in a CDM model at
$z=20$
suffices to remove any discrepancy between the nucleosynthesis range for
$\Omega_b$ and $n=1$ COBE--normalized CDM \cite{SugSilVit}.

Early reionization results in
another possibility that has interesting observational consequences.  Since the
reionization is a consequence of the early non--linearity of very rare dwarf
galaxies, one can well imagine that there may be non--Gaussian smoothing of the
primordial Gaussian fluctuations.  This could arise as follows.  The first
objects to form, in the extreme tail of the distribution, ionize their
environments.  Suppose that this local reionization induces a feedback effect.
For
example, reionization can actually {\it enhance}, ({\it e.g.} \cite{KanSha})
the formation of $H_2$
molecules that form via $H^-$ ions.  The enhanced cooling and gas dissipation
would enhance dwarf galaxy formation with this positive feedback.  One can
imagine this process bootstrapping, much as in the early models of explosive
galaxy formation \cite{OstCow,Ike} that are now ruled out by the low limits on
the Compton $y-$paramete
new last scattering surface and reionization would not occurr simultaneously in
cosmic time.  The reionization process would be patchy and non--Gaussian.  This
could well result in variable optical depth or thickness of the last scattering
surface over degree scales.

\section{Summary}

There is no doubt that early reionization is an essential feature of the
observed Universe.  It
certainly occurred by $z\sim 5,$ and most likely occurred before $z\sim 20$ in
unbiased CDM models.  Diminution of ${\delta T \over T}$ on degree scales by a
factor $\simlt 2$ is almost inevitable in COBE--normalized CDM.  Surviving
fluctuations are plausibly non--Gaussian, since the reionization physics of
rare objects could well introduce local feedback.  A consequence is that
reionization and rescattering generate potentially observable secondary
fluctuations on arc--minute scales.  In CDM models, these secondary
fluctuations are ${{\delta T}\over T} \sim 10^{-6}.$  However in BDM models,
they must be of order $10^{-5},$ and already provide a strong constraint on the
allowable BDM parameter space.

\vskip 1truecm

\noindent I thank E. Bunn, W. Hu, D. Scott, N. Sugiyama and M. Tegmark for
valuable discussions.

\end{document}